# Rate Adaptive Autoencoder-based Geometric Constellation Shaping

**Ognjen Jovanovic, Metodi P. Yankov, Francesco Da Ros and Darko Zibar**
*Department of Electrical and Photonics Engineering, Technical University of Denmark, Kgs. Lyngby, 2800, Denmark*
*ognjo@dtu.dk*

**Abstract:** An autoencoder is used to optimize bit-to-symbol mappings for geometric constellation shaping. The mappings allow for net rate adaptivity without additional hardware complexity, while achieving up to 300km of transmission distance compared to uniform QAM. © 2023 The Author(s)

## 1. Introduction

State of the art coherent optical communications need to be deployed in dynamic network scenarios, which require a certain degree of adaptivity to varying channel conditions [1]. Classically, this is handled by varying the modulation format size, which 1) often produces a coarse granularity in the rate with steps of 2 bits/symbol; and 2) requires the transceiver to support bit to symbol mapping and demapping to/from constellations of different size, increasing the complexity. The probabilistic amplitude shaping (PAS) scheme [1] has emerged as an efficient architecture that provides a solution to the first problem, at the cost of rate matcher and dematcher, which increases complexity. Autoencoders (AEs) are becoming a popular tool to optimize the signaling constellation of digital communication transceivers [2, 3]. In optical communications, AEs have been employed for geometric constellation shaping (GCS) [4], bit labeling [5], mostly with the target of mitigating the impact of fiber nonlinearities, as well as transceiver impairment mitigation [6, 7]. GCS is beneficial over PAS because it does not require explicit matcher and dematcher blocks. However, rate adaptivity with GCS typically requires a rate-flexible forward error correction (FEC), which may increase the complexity of the digital logic. Further, GCS does not allow for straight-forward Gray labeling to be performed, which may lead to sub-optimality when combined with conventional bit-metric decoders as in standard coherent optical communications [8].

In this paper, an AE is used to 1) find optimal bit mappings; and 2) find optimized constellation points for a variety of net rates while maintaining a fixed FEC and demapper logic (i.e. log-likelihood ratio (LLR) computation). The system therefore allows for shaping gain to be achieved with a finer granularity without a complexity increase w.r.t. conventional bit-interleaved coded modulation (BICM).

## 2. System description

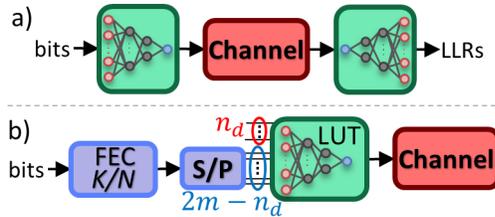

Fig. 1. a) Training setup; b) Testing setup with dummy bit insertion.

The AE-based GCS training setup is given in Fig. 1 a). The mapping function is learned using the AE architecture from [5] to jointly optimize bit labeling and constellation position on the I/Q plane. During the AE training, the transmitter and receiver employ neural networks for mapping of bits and demapping to LLRs, respectively. The transmitter of the proposed system is given in Fig 1. b). During testing, the mapping and demapping functionalities are replaced by a look-up table (LUT) and conventional Gaussian bit-metric receiver [8], respectively.

The AE requires that the modulation format size is known and fixed. For a fixed modulation format size, the generalized mutual information (GMI) typically is penalized w.r.t. the mutual information, as the signal to noise ratio (SNR) decreases. This is due to the penalty in the demapper function related to the inability to resolve constellation points with similar likelihoods at the receiver. An AE implicitly addresses this problem by 'merging' such points closer together and assigning (more than 2) labels with a very small Hamming distance to virtually the same point [7], i.e. a many-to-one mapping (MOM) of bits-to-symbols is produced. Here, we exploit this fact to achieve rate adaptivity in the following way. The GMI of the MOM is analyzed and the bit levels which are ambiguous are not used for data. Instead, they are assigned dummy bits in order to maintain the bit flow and logic at the transmitter and receiver. For a system with an FEC rate of $R = K/N$, a code length of $N$, information block length of $K$ and a modulation format size of $M$, the net data rate per dual polarization channel may be calculated as $R' = (2m - n_d)R$, where $m = log_2 M$ is the number of bit carried by the constellation and $n_d$ is the number of dummy bits in the labeling. If $n_d$ is even, each polarization gets the same number of dummy bits, whereas if it is odd, the allocation is done such that the GMI per polarization is similar for both polarization. In this paper, the $n_d$ bit positions are selected by sorting their per-bit GMI and choosing as many as required from the lowest ones. For example, when the SNR decreases, $n_d$ can be increased. The performance degradation of the large-size constellation at low SNR is compensated for by the increased Euclidean distance of the effectively smaller constellation achieved via MOM. The receiver architecture does not change since K, N, and M are fixed. The only addition w.r.t. BICM is the additional LUTs for mapping depending on the channel conditions.

### 3. Results

A wavelength division multiplexing system is optimized using the nonlinear interference noise model from [9] with dual polarization, 5 channels, 100km spacing, FEC of rate 3/4 and modulation size $M = 256$. Fig. 2 a) shows the maximum distance for a given data rate at which the GMI is above the target rate for the AE-based MOM and uniform QAM (red dotted line, $n_d = 0$) for the central channel. The AE-optimized MOM achieve an extra span of transmission distance with respect to both 256QAM and 128QAM. Here, due to the SNR reduction with the increase of distance, the AE learns a MOM allowing the system to be rate adaptive through insertion of dummy bits without losing the shaping gain or changing the modulation format. Fig. 2 b) shows the constellation learned for 8 spans transmission which does not achieve MOM. In Fig. 2 c), the constellation learned for 20 spans that achieves MOM is shown. In the latter, the AE "merged" some of the points together as shown in Fig. 2 d). The red box shows the bits that are not shared by the 4 points. These 2 bits effectively carry no information and can be assigned dummy bits. The system then exploits the resulting increased Euclidean distance of the constellation to improve the performance.

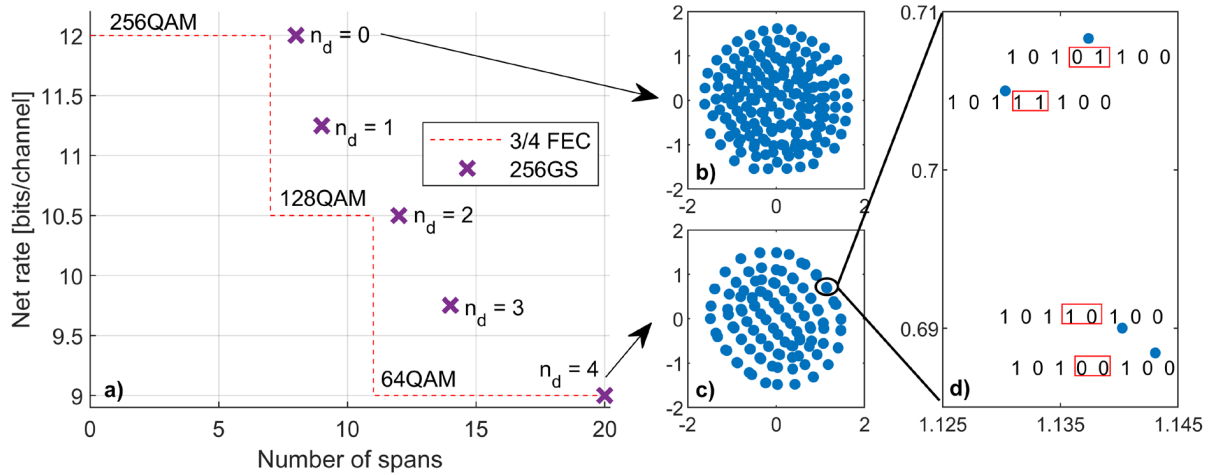

Fig. 2. a) Net rate per dual polarization channel w.r.t. distance for uniform QAM and AE-based GCS with 3/4 FEC; GCS learned at: b) 8 spans; c) 20 spans; d) Zoomed in point to show that 4 points collapsed to each other.

### 4. Conclusion

An autoencoder (AE) is used to optimize rate dependent bit-to-symbol mappings for QAM of fixed size and FEC of fixed rate. Rate adaptivity is achieved through the AE-optimized mapping functions as a result of many-to-one mapping. It achieves shaping gain for a variety of net rates without changing the receiver architecture, the modulation format size or requiring a distribution matcher and dematcher, resulting in a hardware friendly flexible architecture.

**Acknowledgement:** This work was financially supported by the ERC-CoG FRECOM project (grant no. 771878), the Villum Young Investigator OPTIC-AI project (grant no. 29334), and DNRF SPOC, DNRF123.